\documentclass[letterpaper,twocolumn,10pt]{article}
\usepackage{usenix,epsfig,endnotes}
\usepackage[letterpaper, top=0.85in, bottom=0.85in, inner=0.85in, outer=0.85in, columnsep=0.3in, foot=.35in]{geometry}
\usepackage{url}
\usepackage{graphicx}
\usepackage[compact]{titlesec}
\usepackage[font=small,bf]{caption}

\usepackage[ps2pdf,
            hyperfigures=false,
            hyperindex=false,
            pdfborder={ 0 0 0 },
            pdftitle={Graffiti Networks: A Subversive, Internet-Scale File Sharing Model},
            pdfauthor={Andrew Pavlo, Ning Shi},
            pdfkeywords={Graffiti Networks, File Sharing, BitTorrent Alternatives, MediaWiki}]{hyperref}

\begin{document}

\date{}

\title{\huge Graffiti Networks:\\A Subversive, Internet-Scale File Sharing Model}


\author{
{\rm Andrew Pavlo}\\
Brown University \\
{\rm \url{pavlo@cs.brown.edu}}
\and
{\rm Ning Shi}\\
Brown University \\
{\rm \url{ning@cs.brown.edu}}
}


\setlength{\abovecaptionskip}{0.08in}
\setlength{\textfloatsep}{0.1in}

\maketitle


\begin{abstract}
The proliferation of peer-to-peer (P2P) file sharing protocols is due to their efficient and scalable methods for data dissemination to numerous users. But many of these networks have no provisions to provide users with long term access to files after the initial interest has diminished, nor are they able to guarantee protection for users from malicious clients that wish to implicate them in incriminating activities. As such, users may turn to supplementary measures for storing and transferring data in P2P systems. We present a new file sharing paradigm, called a \textit{Graffiti Network}, which allows peers to harness the potentially unlimited storage of the Internet as a third-party intermediary. Our key contributions in this paper are (1) an overview of a distributed system based on this new threat model and (2) a measurement of its viability through a one-year deployment study using a popular web-publishing platform. The results of this experiment motivate a discussion about the challenges of mitigating this type of file sharing in a hostile network environment and how web site operators can protect their resources.
\end{abstract}

\section{Introduction}
In just a few years since its inception, the BitTorrent protocol and similar systems have become the predominant P2P file sharing model~\cite{bittorrent}. But the recent activities of those seeking to take down P2P infrastructures have forced the file sharing community to adapt to a hostile environment~\cite{isps}. Operators of global BitTorrent trackers now take two notable measures in order to indemnify themselves from legal action: (1) the trackers are located in countries that are not party to international copyright treaties, and (2) access to trackers is controlled by private, invite-only communities with strict membership requirements~\cite{bittorrent-communities}. The former allows operators to ignore legal threats to shutdown their services that a law-abiding ISP would normally have to comply with. But this approach can be both prohibitively expensive and difficult to arrange. Additionally, limiting access to only privileged users only temporarily protects a site that has been made private; it takes only a single seditious user to undermine the network and provide damaging evidence to the right parties.

The aforementioned measures may protect tracker operators but they provide little protection to the average file sharing user. This is because the fundamental principle of the BitTorrent protocol is that users download and upload data directly with other untrusted users, rather than download from a single, central source~\cite{bittorrent}. Although some P2P clients employ communication encryption and protocol obfuscation enhancements, such measures do not protect a user from malicious clients that harvest file sharing activity information for future litigation. Furthermore, it has been shown that while it may not be possible to easily view encrypted packet contents, a third-party observer can still deduce that file sharing is occurring by identifying network pairs based on a tracker's public peer list~\cite{sandvine,isps}.

Another limitation of current BitTorrent-like models is that the networks rely on altruistic users to keep files available for others. This is problematic in an environment where users want to limit their exposure to any traffic logging clients, and thus it is in their interest to disconnect immediately once they have the successfully downloaded their desired files. Content in these networks is unavailable once all of the peers that have the complete file depart. Newly arriving clients may be able to download and share some fraction of the data (if any is available), but they must wait and hope that a client returns to the network with the rest of file. Enhancements to private trackers, such as upload/download ratios, provide incentives for clients to continue to seed files~\cite{bittorrent-communities}, but these economic models are difficult to initiate and do little to maintain less popular older files.

In response to the lack of user anonymity and long-term data persistence in existing P2P systems, some users may seek an alternative. But because traditional data hosting solutions are not a viable option for sharing certain content that may have legal consequences, these users must use more questionable means for sharing data. Motivated by this, we developed the \textit{Graffiti Network} distributed file sharing protocol that uses multiple third-party storage sites as a data replication and transfer medium between clients. The Graffiti approach is to use publically available web sites to store multiple copies of shared content. We use the term \textit{graffiti} for our work since we are storing data in a way that non-network participants may regard as unsightly or unwanted vandalism. Our approach presents several new security challenges over other existing P2P systems where clients transmit data directly with each other: (1) a newly arriving peer can still download files even if all other peers have long disconnected, (2) a peer does not need to know about the existence of other peers, and (3) a tracker does not need multiple peers to enforce tit-for-tat policies~\cite{bittorrent}.

The layout of this paper is as follows. First, we provide an overview of the Graffiti Network file sharing model. We then discuss our experimental prototype of the Graffiti Network model that is integrated with a BitTorrent system. The results from our one-year study on the efficacy of our prototype in a real-world deployment show that the use of public storage sites in a file sharing system is possible. We then conclude with a discussion about how both administrators and software developers can guard against such a threat.

\section{Related Work}
\label{sec:related}
We motivate our work by first discussing the related background research and literature.

\subsection{BitTorrent}
\label{sec:related-bittorrent}
The BitTorrent protocol defines the operations of a P2P network that facilitates the efficient sharing of files in a distributed manner~\cite{bittorrent}. Our model inherits many of the features of BitTorrent, but employs third-party storage sites as an intermediary for data transfers, rather than allowing clients to directly download files from each other. This indirection makes it difficult to discover the identities of users that are participating in a Graffiti Network.

The overall efficiency and throughput of BitTorrent systems has been shown to scale gracefully to accommodate many users arriving at the same time to download new and popular files~\cite{bittorrent-modeling}. But while the model works well in the short term, it does not ensure the long term availability of esoteric content or files that become less popular over time. This problem is especially prevalent for content that is released in ``episodes'': new content is shared profusely when it is released, but the number of peers decreases as the file becomes older and newer episodes are released. In a five month study of BitTorrent network activity, it was shown that the average time that a client stays in the network to continue sharing a file after it has received the entire file set was only seven hours~\cite{bittorrent-lifetime}. These results, however, are based on the sharing activity of copyright-free files, and therefore the clients do not have a vested interest in disconnecting immediately. In contrast, a study \cite{bittorrent-measure2} explicitly focused on illegal file sharing activity showed that the departure rate of peers is much faster than previously assumed in \cite{bittorrent-modeling}. The results in \cite{bittorrent-measure1} show that the average availability of a torrent is less than nine days and that most swarms completely die out in only 13 days. Thus, without the incentives for sharing found in private communities~\cite{bittorrent-communities}, most BitTorrent content becomes unavailable after just a short amount of time. To overcome the capricious nature of users, Graffiti Networks use storage sites that have the potential to always be available, and thus the shared files are still accessible after the initial interest in the content has subsided. With enough replication, enforced by a strict asynchronous tit-for-tat model, we believe that a Graffiti Network could provide clients with access to files months or years after it was first introduced to the Internet.

\subsection{Peer-to-Peer Storage Systems}
\label{sec:related-p2pstorage}
Much of the previous work on developing P2P storage systems that provide block storage across multiple nodes is based on distributed hash tables~\cite{cfs,oceanstore,past}. These approaches have the same deficiencies as the BitTorrent model: peers download file blocks directly from other peers, thereby losing anonymity, and the systems do not provide mechanisms to provide long term availability for less popular files after peers disconnect from the network. Other systems are focused on providing anonymous and secure P2P data storage~\cite{publius}. The POTSHARDS system provides secure long-term data storage when the content originator no longer exists using secret splitting and data reconstruction techniques to handle partial losses~\cite{potshards}; their approach assumes multiple, semi-reliable storage backends that are willing to host a client's data. The Freenet anonymous storage system uses key-based routing to locate files stored on remote peers~\cite{freenet}. As discussed in \cite{cfs}, Freenet's anonymity limits both its reliability and performance: files are not associated with any predictable server, and thus unpopular content may disappear since no one is responsible for maintaining replicas.

\subsection{Steganographic Storage Systems}
\label{sec:related-stegstorage}
Although the Graffiti Network model is not a pure steganographic-based storage system, it does share similar properties of this class of systems~\cite{stegvault,mnemosyne}. The Mnemosyne storage service applies the steganography techniques from a local storage system \cite{stegfs} to a distributed hash table~\cite{mnemosyne}. The StegVault proposal uses secret sharing to build a secure P2P storage system on top of reliable multicast~\cite{stegvault}. One key benefit of these systems is that users have plausible deniability of the existence of hidden data because it is concealed inside covering data~\cite{steghide}.

\begin{figure*}[t]
   \centering
   \includegraphics[width=.75\textwidth]{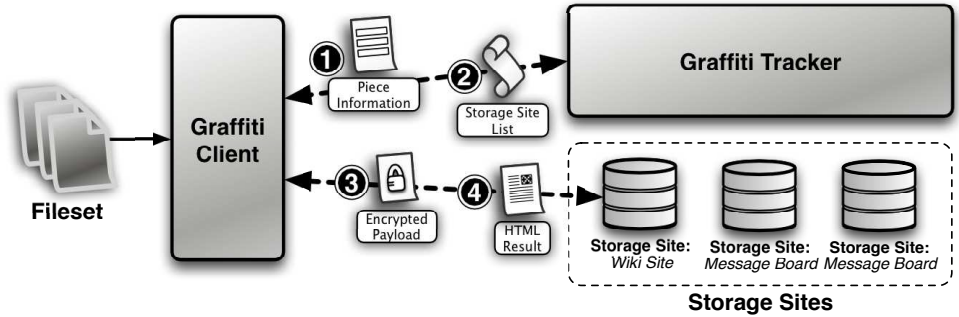}
   \caption{For a given a fileset, the client communicates with the tracker in the following manner: \textbf{(1)} the client sends the tracker the list of pieces it already has; \textbf{(2)} the tracker responds a list of instructions on where the client should download a sub-piece and the location of where to upload a replica; \textbf{(3)} after downloading the new sub-piece, the client then navigates the target storage site and uploads a new encrypted and encoded sub-piece payload; \textbf{(4)} the storage site returns an HTML page and the client verifies that the upload was successful. This process repeats until the client has all the pieces of the fileset and has produced enough replicas for the tracker.}
   \label{fig:overview}
\end{figure*}

\subsection{Alternative Storage Sites}
\label{sec:related-storagesites}
Since the Graffiti Network model relies on gaining access to and the circumvention of third-party storage sites to host content, we consider the alternative approach of using dedicated storage services that are explicitly designed for the storage and transfer of large files. The Amazon Simple Storage Service provides a well-defined API for writing arbitrary data files, but it currently charges for both the storage space an account uses as well as the network bandwidth used to transfer data~\cite{amazon-s3}. The Gmail Filesystem enables Google email accounts to be used as a network storage medium, but adopting approach would require users to share account information~\cite{gmail-fs}. The Usenet news service is another potential storage system, but servers often impose a message retention time and many ISPs have discontinued providing this service to customers for free.

Free web-based file-hosting sites also do not provide the robustness that we seek in our file sharing model~\cite{rapidshare}. One limitation of these sites is that large files are broken into separate downloads and users must wait for some time period before they are allowed to retrieve the next piece. Furthermore, the user must manually enter each segment URL into their browser and repeatedly pass human-validation tests~\cite{captcha}. These free hosting sites are also under scrutiny because many of their users post illegal content, and thus the site operators streamline the removal process for files and the disclosure of offending users' information for copyright holders in order to quickly diffuse any legal action that may disrupt the hosting site's revenue stream. Despite this, it is possible to include file-hosting sites as just one of the many options available in a Graffiti Network deployment (see Section~\ref{sec:sites}).

Lastly, another proposed solution is to create a highly-volatile storage site by sending data packets to unsuspecting network entities to leverage network latency as a type of durability~\cite{juggling}. The idea is to continuously send data to targets that relay the same data back to the source, therefore two copies of the data are always theoretically available. This approach is not practical for the Graffiti Network model because it does not allow the data to be shared amongst multiple peers. Furthermore, it requires that the original data source remain online in order to keep cycling the packets back out over the wire.

\section{Graffiti Network Model}
\label{sec:model}
We now describe how a file-sharing system based on the Graffiti Network model would operate. We discuss various measures and techniques that ensure the system is stable, usable, and scalable. Such qualities are necessary to facilitate wide-spread adoption by file-sharing participants, thereby making the threat a real possibility.

To describe the Graffiti model, we adopt the terminology of the BitTorrent protocol~\cite{bittorrent}. We define a \textit{fileset} as a set of one or more files that peers wish to share. The fileset's data is divided into multiple fixed-length \textit{pieces} of $n$ bytes (the last piece can contain less than $n$ bytes) and are numbered sequentially. Each piece is divided further into fixed-length \textit{sub-pieces}. A Graffiti Network that is deployed to distribute these pieces is comprised of three distinct components: (1) a \textit{tracker} coordinates the replication and sharing procedures of a fileset, (2) a \textit{client} downloads and replicates the fileset data managed by the tracker, and (3) third-party storage \textit{sites} store and provide access to fileset data for peers. Any client that wishes to download and reconstruct the original fileset is required by the tracker to produce multiple sub-piece \textit{replicas} on as many storage sites as possible.

A high-level overview of the Graffiti Network protocol is shown in Figure~\ref{fig:overview}. To connect to the Graffiti Network, the client first announces itself to the tracker and provides it with a list of all the pieces that the peer has already downloaded. The tracker responds with a series of sub-piece \textit{request pairs} for a new piece that the client is missing. Each request pair consists of (1) a download location where the peer can retrieve a sub-piece and (2) instructions to produce a new replica on a different storage site for the data it just downloaded. Graffiti trackers follow a strict tit-for-tat protocol: for each sub-piece that a peer downloads, that peer is required to generate a replica for a previously downloaded sub-piece on a different storage site and send the location of this new replica back to the tracker before it can receive the next piece.

\subsection{Central Tracker}
\label{sec:tracker}
The tracker provides a directory service for peers to retrieve a fileset. For each piece of data in a fileset, the tracker maintains a table of the sub-piece replica locations on sites that were generated by clients. Each replica is annotated with three pieces of meta-data: (1) a unique encryption key for that replica, (2) a checksum for each sub-piece, and (3) the first and last byte sequences of the encrypted data block on the storage site. The tracker uses a different encryption key per entry to ensure that each replica is stored as a unique character sequence to prevent the use of tools to discover other replicas. The checksum and sequence markers also allow peers to determine whether a replica has the proper byte sequence and to locate data boundaries at the storage site location.

For each connected peer, the tracker maintains an \textit{active piece set} (APS) of download/upload replica pairs that are unfulfilled requests for a client. Each pair consists of a sub-piece identifier that the tracker provided for client to download and a storage site location where the tracker instructed the client to make a new replica. Once the client provides the tracker with information about a new replica for a download/upload pair, the entry is removed from that client's APS and the client is allowed to receive new information. The size of the APS is determined by the tracker's administrator and prevents a client for downloading too many sub-pieces without producing any new replicas. As in the BitTorrent protocol, the Graffiti tracker strives for uniform availability of all data pieces~\cite{bittorrent}. Since the tracker decrees what pieces the clients must replicate for each request in the APS, it can decide to replicate the ``rarest'' pieces first.

Malicious clients in Graffiti Networks are quite different than malicious clients in BitTorrent networks~\cite{bitthief}. A rogue Graffiti client may have other ulterior goals: (1) to discover all of the storage site locations used by a tracker in order to contact site administrators and have the replica data removed or (2) to falsely identify valid storage sites and replica locations as invalid in an attempt to disrupt operations. In the first case of trying to discover all of a fileset's replicas, the tracker can use throttling measures to prevent a client from learning too much in a short amount of time. But for the latter problem, the tracker should not actively check whether a client actually uploaded the data at the location it claims it did, due to security and economic reasons. Instead it can employ proxies or other third-party entities to determine whether a client is behaving properly. For example, the tracker can retrieve a page through the Coral Cache or Tor services to determine if the data was stored at the location claimed by a client~\cite{freedman04,dingledine04}.



\subsection{Client}
\label{sec:client}
A Graffiti client allows a user to automatically download a fileset stored on one or more storage sites. A user must first obtain a metadata file for a specific fileset uniquely identified by an ``info hash'' in order to begin downloading~\cite{bittorrent}.

After the client first announces itself to the tracker at the address listed in the metadata file, the tracker places the peer in an ``initialization'' mode. This is always done regardless of whether the client is connecting for the first time or if it is returning with some pieces already downloaded. The tracker sends every new client the same \textit{initial piece set} (IPS) that will use for the first phase of downloading and replication. This initial set is the same for all clients arriving within a certain time period to prevent a client from initiating multiple new connections without ever creating new replicas. The size of the initial set is the same size as the APS and its information is changed to a different random set of sub-pieces at regular intervals (e.g., hours or days, rather than minutes). Thus, it is possible for a rogue client to retrieve a complete fileset without ever producing a new replica for the network, but it would take several days or weeks to cycle through all of the tracker's IPS combinations if there were a significantly large number of pieces. The client is required to also produce two new replicas for each sub-piece in the IPS, even if the client has already downloaded the pieces previously. This policy is akin to a new tenant paying ``last month's rent'' before moving into an apartment: it ensures that client cannot disconnect from the network without creating new replicas for each piece that it downloads.

Once the client successfully downloads and generates sufficient replicas for its IPS, it leaves the initialization phase and is then allowed to receive arbitrary pieces. The protocol works the same before: the tracker maintains an APS for each client and only gives new download locations once that particular client has produced a new replica on a storage site.

\subsection{Storage Sites}
\label{sec:sites}
A potential Graffiti storage site is any accessible network entity that allows for data to be stored and retrieved using a known network protocol. In practice, peers will likely use publically available web sites that provide services that Graffiti clients repurpose to store arbitrary blocks of data. This approach has the distinction that all data movement appears as normal HTTP traffic, and thus is immune to current ISP throttling and tracking techniques~\cite{isps}.

The ideal storage site for a Graffiti Network is one that allows for anyone to post data without CAPTCHA protections~\cite{captcha} and is either unmoderated or has long abandoned by its owner. A popular and high-traffic wiki site, for example, would not be a good storage site candidate as it likely that non-malicious visitors would quickly notice the changes made by Graffiti clients to store replica data. With the rise of many open-source web-publishing platforms, there are many potential targets that allow for anonymous or semi-anonymous data posting. Notable examples include paste-bins, wiki sites, message boards, and blogs. An HTML-based storage site also allows the data to be disseminated to peers through disparate channels once it is online, such as through Coral Cache~\cite{freedman04} or Tor~\cite{dingledine04}. The data embedded in the site's pages could also be picked up by search engine caching and archiving services for longer-term storage.

Other potential storage sites include any photo and file hosting sites that allow for automated data uploading. In the case of the former, the data could also be hidden inside of image files using well-known techniques~\cite{steganography,ramkumar01}. As the Internet evolves, new targets will emerge that can be incorporated into existing networks. The system could also allow clients to use storage sites that are password protected for writing data, but where an account is not required to read back the data. This obviates the need for a client to send the tracker account information, which could then be used improperly by other clients to tamper with or destroy the data.

Using involuntary web sites as storage dumps seems counterintuitive if the main goal of the network is data persistence and availability, since replicas are promptly removed when site administrators and moderators discover them. The Graffiti model overcomes this challenge and takes advantage of ``free network storage'' through a massive replication and obfuscation process. It is not trivial, however, to automatically store arbitrary data on random web sites nor is it trivial to discover which sites are available with the properties stated above. The prevalence of popular web publishing software means that one only needs to target a small number of platforms in order to circumvent a large portion of the Internet. Furthermore, many sites, such as wikis and message boards, often display the network location of the user responsible for adding new content or making changes to their pages, which makes it difficult to deny responsibility for participating in illegal activities. We argue that by fracturing a fileset's replicas across hundreds of storage sites, it is difficult to be fully implicated when only a fraction of the evidence is available. A distributed effort to probe websites and uncover open storage paths could allow peers to draw on a nearly limitless pool of available storage.

\section{Experimental Deployment}
\label{sec:simulation}
To determine whether the Graffiti Network model is a viable and thus is a potential threat, we implemented a prototype Graffiti tracker and client as an extension to the BitTorrent protocol. We then stored a sample data set on a large number of open sites and measured the availability of our data for almost an entire year.

We built our system on top of the open-source libtorrent~\cite{libtorrent} BitTorrent library in order to allow clients to participate in torrent swarms concurrently with Graffiti Network activities. When enough peers are available, the client operates strictly in BitTorrent mode. But if the number of distributed copies in the swarm drops below a threshold, the client begins to contact the tracker using the Graffiti protocol in conjunction with its BitTorrent operations. As new pieces are retrieved from storage sites, they are passed to libtorrent's storage manager for seeding to other peers.

\subsection{Storage Site Discovery}
In our experimental prototype, we target the open source MediaWiki~\cite{mediawiki} platform as the potential storage site for the network. Due to the popularity of sites like Wikipedia that use MediaWiki, we believe that it is the most widely deployed wiki platform with a large number of less-experienced users that install the software without changing the permissive default settings. Another key characteristic is that the MediaWiki platform maintains a complete revision log for each article, which allows Graffiti peers to retrieve data even if the changes are reversed or the content is altered.

We decided to test our system on open MediaWiki sites that we do not have control over as this allows us to best measure whether our assumptions about how long the data will remain on the sites are correct. We developed a distributed web crawler to discover MediaWiki installations through search engines using keywords that are uniquely indicative of a newly installed site. The crawler purposely ignored well-known MediaWiki sites (e.g., those sites that are part of the Wikipedia Foundation) and the commercialized versions of MediaWiki (e.g., Wikia). For each site that the crawler found, we probed it to determine what kind of protection scheme it utilizes and the last time that it was updated (see Table~\ref{tab:sites}). Of the 23,156 unique MediaWiki installations that we found, 8,483 sites allowed for anonymous editing and 5,983 allowed users to register accounts without CAPTCHA or email protections in order to make edits~\cite{captcha}. The default MediaWiki installation provides a primitive arithmetic ``puzzle'' protection countermeasure that we found in use on 1,157 sites; this puzzle is easily broken with just a few lines of code, and thus did not prevent our system from storing data on these sites. Lastly, in order to minimize the impact of our experiments, we only targeted those sites that had not been updated within the last three months, thereby reducing our list to 5,646 sites; lowering the threshold to two months would have yielded a total of 11,987 potential storage sites.

\begin{table}[!t]
   \centering
   \begin{tabular}{lrr}
      & Sites Found & Sites Used \\
      Anonymous Edits & 8,483 & 3,161 \\
      Registration Protected & 5,983 & 2,347 \\
      Puzzle Protected & 1,157 & 138 \\
      CAPTCHA Protected & 1,586 & - \\
      Not Publicly Modifiable & 5,946 & - \\
      \hline
      \textbf{Total:} & \textbf{23,156} & \textbf{5,646} \\
   \end{tabular}
   \caption{The categories of protection used by the MediaWiki sites discovered during the collection process and the sites used in the experimental deployment.}
   \label{tab:sites}
\end{table}

\begin{figure*}[!ht]
   \centering
   \begin{minipage}{3.2in}
      \centering
      \includegraphics[width=3.25in]{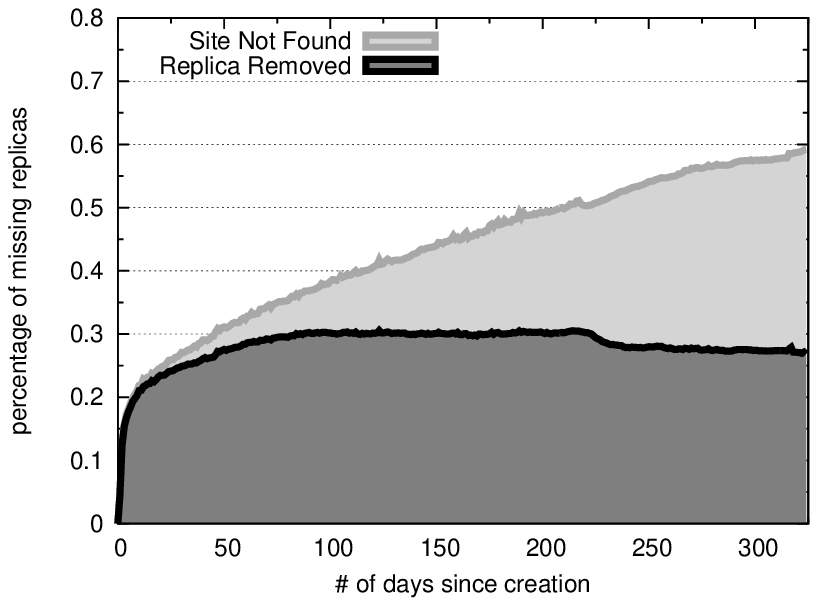}
      \caption{Percentage of total replicas removed over time categorized by the type of failure.}
      \label{fig:graph-visits}
   \end{minipage}
   \begin{minipage}{0.3in}
      \hspace*{0.3in}
   \end{minipage}
   \begin{minipage}{3.2in}
      \centering
      \includegraphics[width=3.2in]{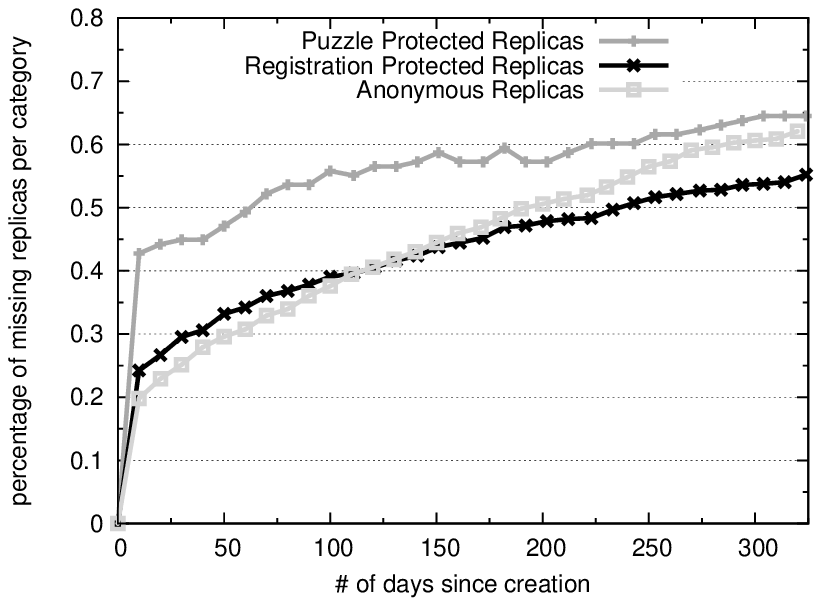}
      \caption{The availability of replicas categorized by its corresponding storage site's protection schemes.}
      \label{fig:graph-availability}
   \end{minipage}
   \vspace{-.15in}
\end{figure*}

The Graffiti client stores data on MediaWiki sites as base64-encoded, Blowfish-encrypted blocks of text that are written in a new article titled with a random word from the dictionary. A more resilient approach would be to modify a popular page on a given site, and then immediately reverse the changes and mark the revision as vandalism. This has two significant implications compared to writing data to a newly created article. Foremost is that removing this data completely from the page's history requires administrators to delete the entire page from the database and restore the latest revision by hand, thereby losing all the previous legitimate revisions. Second, such an attack is more likely to be overlooked by a site's operators since they may only care whether the changes were reversed. We deemed this technique too malevolent for the purpose of our experiments, and thus chose to not implement it.

To retrieve a sub-piece stored on one of these storage sites, the client downloads the web page and extracts the text surrounded by the byte sequence markers provided by the tracker. The client then reverses the base64 encoding, decrypts the data, and verifies that it matches the checksum provided by the tracker.

\subsection{System Configuration}
For our experimental deployment, we used a Linux ISO split into 512KB pieces and 64KB sub-pieces as our sample data file that the clients want to share. Even though we were able to store up to 512KB payloads on a single MediaWiki page, we choose to use a smaller sub-piece size. Again, another more malicious approach would be to store a payload with the size that can be uploaded and retrieved but causes either a browser or the server to choke if the operator tries to access the page through the MediaWiki administrative interface. For example, we found that it was possible to store 512KB pieces that would exhaust the default 20MB memory limit of PHP if someone tried to remove the data. Thus, the only way to remove the content is to execute the proper SQL commands directly in the database, which is likely too difficult for most users.

We initiated file sharing activity on April 10th, 2009 using a tracker and five clients deployed in our departmental lab. Each client connects to the tracker and produces a full copy of a sub-piece on one of the 5,600+ MediaWiki sites. We assume that all clients are truthful about whether a replica is available and do not falsify replica URLs. We instrumented the tracker to target each storage site only once (although variations in sub-domains and URL rewriting led to some sites being used more than once).

Along with the data payload, at the top of each wiki page we stored a small paragraph with an explanation of the seemingly random text. This description also included a unique tracking link back to our web page with further information about the project. Tracking users' click-throughs from these links allows us to measure to some extent whether humans were actually discovering our payload pages before they were deleted.

Once the clients pushed out all of the data to the sites, we then used a separate tool to check daily whether the data we stored is still in place and has not been modified. We check every replica regardless if it has not been available for some time to ensure that the errors are not transient.

\begin{figure}[t]
   \centering
   \includegraphics[width=3.2in]{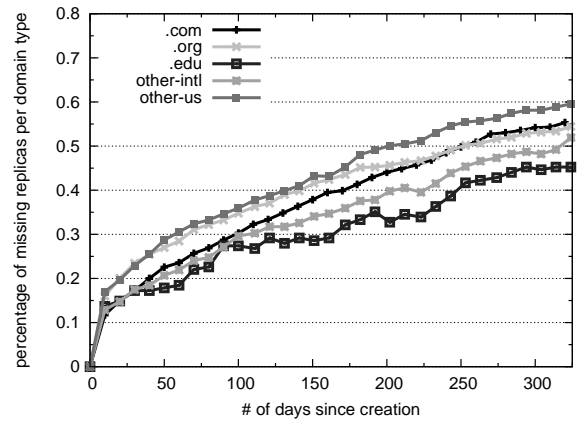}
   \caption{The cumulative availability of replicas categorized by their domain type: .com (42.5\%), .edu (3.2\%), .org (24.1\%), US-based other (14.0\%), and Non-US-based other (16.1\%). }
   \label{fig:graph-domain}
\end{figure}

\subsection{Results}
\label{sec:experiments-results}
We now report on the availability of the 5,646 replicas that we stored in our experiments from April 10th to February 28th, 2010. For each missing replica, we categorize the replica as (1) \textit{removed} if the site is available but the original page is missing, (2) \textit{changed} if both the site and the original page are available, but the data does not match our stored checksum, or (3) \textit{not found} if the site is no longer available (e.g., the domain name has expired or MediaWiki was uninstalled). Our investigation found that the missing replicas were only either removed or not found; no replica had its contents altered.

On the last day of our data collection, roughly 40\% of the replicas were still available and hosting the original data that the prototype clients uploaded. The graph in Figure~\ref{fig:graph-visits} shows a timeline of the percentage of replicas that are not available on each day that we checked. The first notable data point is that an initial 20\% of the replicas were removed within the same week that they were created. The rate in which sites are removed then tapers off as time progresses. We attribute this drop-off in activity to two possible reasons. Foremost is that by default any changes to a MediaWiki site will appear on the first page of revision logs for seven days after the revision is created, and thus our actions are more likely to be discovered soon after the data is posted. The second possible reason is because a story about our project appeared on the front page of a popular technology news website on the third day of our experiment~\cite{slashdot}. We believe that the ``notoriety'' of the project during this period may have caused administrators to examine their websites to see if they were targeted by our system. Once this initial attention diminished, the slopes of the lines in Figure~\ref{fig:graph-visits} decrease and it takes another 35 days before another 10\% of the replicas are removed. After about 100 days, the growth rate of replicas being removed (i.e., the lower portion of the curve in Figure~\ref{fig:graph-visits}) tapers off and the number of sites that become unavailable begins to rise. This is expected since many of the sites were not actively used by their proprietor, and thus are taken down arbitrarily.

The graph in Figure~\ref{fig:graph-availability} shows how the replicas were removed over time in relation to their storage site's protection scheme. The salient aspect of the result is that initially sites that employed some type of protection were faster to remove replicas. This is expected, since many of the sites that employed some protection were still being used by users despite having not been updated recently, whereas many of the completely open sites still displayed the default MediaWiki homepage message and thus were never even used once they were installed. Such sites are likely long forgotten by their owners who may never discover the replicas once they pass the default seven day revision log window. But after approximately 120 days, the percentage of missing replicas stored on sites allowing for anonymous edits surpasses sites using the basic registration protection.

Lastly, the graph in Figure~\ref{fig:graph-domain} charts the availability of replicas with respect to the domain name of the storage site. We attribute greater durability of data stored on .edu and .org sites compared to other domains; such organizations are likely to use open-source software for collaboration and internal sites are often not behind corporate firewalls.

\section{Discussion}
\label{sec:discussion}
The results presented in the previous section clearly demonstrate the efficacy of the Graffiti Network model as a means for facilitating longer-term file sharing. We therefore argue that the threat of such a system does indeed exist and sites need to take measures to protect themselves from being used in such a manner that we have describe.

\subsection{Countermeasures}
Much of the feedback that we received on the project was from administrators that expressed their desire to provide an open wiki site that allowed anonymous contributions, despite the inevitable exposure to vandalism and spam. We counter that such sites that do not want to require users to register an account should still use CAPTCHA protections, such as before a user is allowed to edit a page. In practice, we found that the \mbox{reCAPTCHA}~\cite{vonahn08} project is the most effective protection as it does not require administrators to install special server-side graphics libraries and strikes a proper balance between availability and complexity. More complex CAPTCHA schemes would not deter future Graffiti clients that are able to solve CAPTCHAs (either manually or programmatically) and may only inhibit legitimate visually impaired users. If sites wish to still remain open, the CAPTCHA could be selectively enabled only when an unverified user tries to post data larger than some low default threshold or creates too many new pages in a short time span.

We also believe that other simple protection measures could be included in popular web applications to prevent abandoned or forgotten sites from being used for unintended purposes. For example, MediaWiki's default behavior could be to lock down the editing features of a site after a certain number of days if it was installed but then never actually used. This approach is similar to the one used by some blogging platforms to disable comments on older posts. Administrators could easily re-enable this functionality by simply logging into the site again. Another technique is to use a page counter that is invoked on the client-side (e.g., through JavaScript) and then compare the results with server-side logs to determine whether there are an unusually large number of users accessing pages through a non-browser client. Web application frameworks, such as Ruby on Rails and Django, could also provide similar features to protect custom-made sites.

\subsection{Variations \& Adaptations}
Other than for P2P activities, the Graffiti model is also of potential use for large-scale distributed systems used by criminal organizations, often referred to as \textit{botnets}. The goal of most botnet operators is to gain access to a large supply of computational resources for purposes of network communication (e.g., sending emails or DOS attacks). If these goals shift towards more data-centric activities, then systems based on some of the principals of the Graffiti Network model may become prevalent in order to store large amounts of data for the botnet. Alternatively, instead of storing replicated data, the commandeered storage sites could also be used as a control channel for other entities in the botnet.



\section{Acknowledgments}
The authors would like to thank Arvid Norberg at BitTorrent, Inc. for his assistance with the libtorrent library~\cite{libtorrent}.

\section{Conclusion}
\label{sec:conclusion}
We have presented an overview of Graffiti Networks, a new file sharing model that allows peers to subversively use third-party storage sites as an intermediary for transferring files between users. Our client-tracker paradigm is similar to the BitTorrent protocol, but is designed to provide long term file availability to users while preserving their anonymity. We do not intend the Graffiti model to supplant BitTorrent networks, as it will never achieve the same maximum network throughput nor will it ever be as efficient. We believe, however, that our approach can have a symbiotic relationship with existing deployments: peers would use a Graffiti Network-like system to improve the long term availability of shared files, while leveraging the faster initial transfer rates of direct P2P communication for data dissemination. We have implemented a prototype and shown that data can be stored on publically accessible sites for extended periods of time, beyond what is often possible in other existing peer-to-peer systems. After almost an entire year, roughly 40\% of the data that we stored on sites that are not under our control was still available. These results indicate that malicious users may adopt the Graffiti Network model, and thus site operators should take measures to prevent their sites from being used in this manner.


{\footnotesize
 \linespread{0.85}
\bibliographystyle{acm}
\bibliography{graffiti}
}

\end{document}